# High Photocurrent in Gated Graphene-Silicon Hybrid Photodiodes

Sarah Riazimehr[a,c], Satender Kataria[a,c,*], Rainer Bornemann[a], Peter Haring Bolivar[a], Francisco Javier Garcia Ruiz[b], Olof Engström[a], Andres Godoy[b], Max C. Lemme[a,c,*]

[a] University of Siegen, School of Science and Technology, Department of Electrical Engineering and Computer Science, Hölderlinstr. 3, 57076 Siegen, Germany

[b] Dpto. de Electrónica y Tecnología de Computadores, Facultad de Ciencias, Universidad de Granada, Av. Fuentenueva S/N, 18071 Granada, Spain

[c] RWTH Aachen University, Faculty of Electrical Engineering and Information Technology, Chair for Electronic Devices, Otto-Blumenthal-Str. 25, 52074 Aachen, Germany

*email: Satender.kataria@rwth-aachen.de, max.lemme@rwth-aachen.de

## Abstract

Graphene/silicon (G/Si) heterojunction based devices have been demonstrated as high responsivity photodetectors that are potentially compatible with semiconductor technology. Such G/Si Schottky junction diodes are typically in parallel with gated G/silicon dioxide ($SiO_2$)/Si areas, where the graphene is contacted. Here, we utilize scanning photocurrent measurements to investigate the spatial distribution and explain the physical origin of photocurrent generation in these devices. We observe distinctly higher photocurrents underneath the isolating region of graphene on $SiO_2$ adjacent to the Schottky junction of G/Si. A certain threshold voltage ($V_T$) is required before this can be observed, and its origins are similar to that of the threshold voltage in metal oxide semiconductor field effect transistors. A physical model serves to explain the large photocurrents underneath $SiO_2$ by the formation of an inversion layer in Si. Our findings contribute to a basic understanding of graphene / semiconductor hybrid devices which, in turn, can help in designing efficient optoelectronic devices and systems based on such 2D/3D heterojunctions.

Graphene-based optoelectronic devices and photodetectors have recently attracted scientific attention for their ultrafast response time and broadband spectral range [1–11]. Graphene is largely compatible with the well-established Silicon (Si) process technology, which makes it a promising candidate for large-scale integration and cost-effective applications [12–14]. Although pure graphene-based photodetectors are extremely attractive for ultrafast optical communications, they suffer from low light absorption and hence low photoresponsivity. During the past few years, there have been many studies to understand the fundamentals of light matter interaction in graphene [15–20], and to improve the photoresponsivity using complex architectures [3,21–26]. Nevertheless, practical applications may require hybrid technologies, i.e. co-integration of graphene and conventional semiconductors, as demonstrated for high-speed communications [27], solar cells [28], chemical and biological sensing [29] and photodetectors [30–33]. Graphene/Si (G/Si) Schottky junctions are among the simplest possible hybrid structures, and substantial experimental and theoretical work has been published on such heterojunctions for photodetection [31,33–39]. Here, the photogeneration of charge carriers mainly occurs in Si due to low absolute light absorption in graphene (2.3%) [7]. Nevertheless, graphene forms a transparent Schottky junction with Si that enables the extraction of photoexcited carries and thus allows exposing the entire active area to the incident photons. Most of the studies to date have focused on improving the performance of G/Si Schottky diodes in terms of photoresponse and on the nature of the Schottky barrier. Less attention has been paid to the exact location of photocurrent generation in G/Si Schottky diodes. Liu *et al.* [40] investigated the role of different substrates on the photoresponse of graphene using scanning photocurrent measurements, and they observed that photocurrent in graphene on insulating $SiO_2$ substrates is larger than photocurrents obtained in graphene on Si substrates. This observation was attributed to the likelihood of carriers excited from the trap states in $SiO_2$ and their much longer lifetimes as compared to those excited in Si. Recently, Srisonphan et al. [37] have reported extremely high quantum efficiencies in hybrid graphene/Si based devices where graphene is placed across $SiO_2$ over a Si trench to form a device consisting of a graphene-Si heterojunction and a graphene/$SiO_2$/Si field effect structure. They propose photoinduced carrier multiplication in the 2DEG region near $SiO_2$/Si interface to explain the observed high photocurrents in their devices. Despite considerable work carried out on graphene/Si heterojunction-based devices, the main mechanisms of the observable high photocurrents in these devices have been only recently investigated in depth [41,42]. In this work, we thoroughly investigate graphene/n-Si Schottky photodiodes using the scanning photocurrent measurement technique. We clearly show where charge carriers are photogenerated and injected from the Si substrate to graphene. Counterintuitively, we find that higher photocurrent is generated in the graphene/$SiO_2$/Si (G/$SiO_2$/Si or GIS) region compared to the graphene/Si (G/Si) region at reverse biases above a threshold voltage ($V_T$). This observation is found to be independent of excitation laser power and it is explained through simulations by the formation of an inversion layer in n-Si under G/$SiO_2$, corroborating the experimental data and the model proposed in [41,42].

The G/Si Schottky diodes were fabricated using chemical vapor deposited graphene, transferred onto pre-patterned n-Si substrates, similar to the process in [33]. Fig. 1a and 1b show a schematic and a scanning electron micrograph of a G/n-Si photodiode, respectively. One end of the structured graphene film is in contact with the Si substrate, forming the Schottky junction. The other end is contacted with a gold pad on $SiO_2$.

Current density-voltage (J-V) characteristics of the G/n-Si Schottky photodiode are shown in Fig. 1c in semi-logarithmic scale. The black and the red plots (Fig. 1c) are representative of J-V characteristics of the diode in the dark and under illumination, respectively. The photodiode clearly exhibits rectifying behavior in the dark. The forward J-V characteristic of the diode can be described by the Schottky diode equation [43]. For this diode, an ideality factor of 1.16, a Schottky barrier height (SBH) of 0.76 eV and a series resistance of 4 kΩ have been extracted (the method is described in detail in [44]). The ideality factor of 1.16 indicates how closely the diode follows an ideal diode behavior with an ideality factor of 1. A detailed discussion on interpretation of these parameters can be found in [39]. The corresponding energy band diagrams for the G/n-Si diode at zero bias voltage in the dark is shown in Fig. 1d. The photoresponsivity of the G/n-Si diodes has been probed under white-light illumination with an intensity of 0.5 mW.cm$^{-2}$. When the G/n-Si junction is illuminated, the incident photons generate electron-hole pairs in the n-Si substrate. Under the application of a reverse bias, the photogenerated holes in n-Si are accelerated into graphene, leading to a significant photocurrent. As a result, the diode in the off-state under reverse bias exhibits a dark current density of 38 µA.cm$^{-2}$, while under illumination a noticeable photocurrent density of 2 mA.cm$^{-2}$ was measured at -2 V. The energy band diagram of the photodiode in reverse bias under illumination is shown in Fig. 1e.

The absolute spectral response (SR) was measured using a lock-in technique by a LabVIEW controlled setup. Fig. 1f shows the SR measurements over a broad spectrum (from 360 nm to 1800 nm) at various applied reverse bias voltages ($V_R$) on the G/n-Si photodiode. The plot shows an increase of the absolute SR value with the applied reverse bias due to the increased electric field (Fig. 1f). The maximum responsivity is 270 mA.W$^{-1}$ at a reverse dc bias of -2 V. This maximum is observed at an energy of approximately 1.30 eV ($\lambda$ = 950 nm) and can be attributed to absorption in the n-Si. In fact, the photodiode shows a SR very similar to Si p-n photodiodes in shape and magnitude, even though one doped Si region has been replaced by single layer graphene. In contrast to Si p-n diodes, a low and flat SR can be observed over a broad spectrum for energies below the Si bandgap, where there is no contribution from the underlying n-Si. In this region, the SR drops to values below 0.19 mA.W$^{-1}$. This part of the SR can be attributed to light absorption of 2.3 % in the single layer graphene, as reported in our previous work [33].

These data establish that the fabricated devices behave as photodiodes similar to our previously published devices [33]. Next, we performed scanning photocurrent (SPC) measurements to map out the regions of photocurrent generation in the devices. It should be remembered here that J-V characteristics or SR of a diode are generally measured by shining light on the whole device area, therefore, it provides an overall picture of the diode behavior and does not distinguish the regions (of photocurrent generation) constituting the device structure. However, SPC measurements can provide a detailed spatial distribution of photocurrent in localized regions by scanning a laser of desired wavelength over a selected area. Fig. 2a shows an optical micrograph of the device under study. The red rectangle indicates the scanned area, and the black dashed line represents the area where graphene is present. The device was imaged over an area of 0.65 mm x 1.3 mm with a scan speed of 0.256 s/line and integration time of 1 ms. We used a 10 × objective (numerical aperture 0.25) for obtaining the large area scans. The diffraction limited laser spot size was approximately 2.4 µm. Fig. 2b, 2c and 2d show the photocurrent maps of the scanned area at $V_R$ = -1 V, -1.5 V and -2 V, respectively, at a low laser power of 2 µW. At a reverse bias of -1 V, higher photocurrents were recorded in the G/Si region compared to the G/SiO$_2$ region (Fig. 2b). This situation was reversed for increased $V_R$, where much higher photocurrents were obtained in the G/SiO$_2$ regions (Fig. 2c and 2d). Comparing the absolute increase of photocurrent with $V_R$, we observe a eight fold rise in current for G/SiO$_2$, and a two fold rise in G/Si region at -2 V compared to -1 V (Fig. 2e). This result is quite counterintuitive to the initial perception that photocurrents are generated predominantly in the G/Si region, i.e. at the Schottky junction, and will be explained below. We confirmed these measurements on another device, which also exhibited good rectification behavior, with similar results for the photocurrent (Fig. S1). We also note a decaying photocurrent near the graphene edge in G/SiO$_2$ and G/Si region (Fig. S2). This can be attributed to the rapid separation of photogenerated charge carriers in Si near these junctions.

Next, we varied the incident laser power (by two orders of magnitude) and performed the SPC measurements on the same area at a constant $V_R$. This was done in order to exclude any dependency, if any, of the present findings on the incident laser power. We selected $V_R$ to be -1 V and -2 V, at which photocurrent was higher in the G/Si and in the G/SiO$_2$ region, respectively. The photocurrent maps of the photodiode at $V_R$ = -1 V with laser powers of 0.5 µW, 5 µW and 50 µW are shown in Fig. 3a to 3c. A slight increase in photocurrent with laser power is noted, but higher photocurrents were consistently measured in the G/Si region. The current maps obtained at $V_R$ = -2 V for different laser powers again show a different picture (Fig. 3d to 3f). Here, the photocurrent was consistently higher in G/SiO$_2$ region than in G/Si region. The average values of the so obtained photocurrents at different laser powers are plotted in Fig. 3g. The maps show a substantially sharper rise in the photocurrent in the G/SiO$_2$ region compared to the G/Si one when V$_R$ was varied from -1 V to -2 V at a defined laser power. We can conclude that the region that produces the higher photocurrent is independent of the incident laser power, instead it depends on the applied bias voltage.

I-V curves of G/SiO$_2$ and G/Si were recorded while shining the laser (used for SPC measurements) locally in these regions one after the other (Fig. 3h). Unlike, the measurements where the whole diode is illuminated with white light to obtain I-V curves, these local measurements yield more detailed information by isolating the photocurrent generation and contribution from the different regions. A kink in I-V curves is observed in the G/SiO$_2$ regions at V$_R$ of 1.2 to 1.3 V, irrespective of the incident laser power, after which the current rises sharply. We deliberately used two laser powers in order to cover the regimes of low (5 µW) and high laser power (150 µW), respectively. These observations clearly demonstrate that G/SiO$_2$ regions play a significant role in PC generation, and that there appears to be a threshold voltage where these areas become more efficient. Our device structure can clearly be considered as a combination of two heterostructures, namely, graphene on n-Si (G/Si) and graphene-SiO$_2$-n-Si (GIS). The first one acts as a Schottky junction and the second one behaves as conductor-insulator-semiconductor.

To further elucidate the physical origin of the observed high photocurrent in the G/SiO$_2$ regions of the photodiodes, we have assessed the evolution of the energy bands as a function of the applied bias, highlighting the different behavior in each region. In the dark, the surface potential in the Schottky diode is equal to $V_{bi} - V_a$, where built-in potential $V_{bi} = (\phi_g - \phi_{Si})/e$. $e$, $\phi_g$, $\phi_{Si}$ and $V_a$ refer to the electron's charge, graphene's work function, Si work function and the diode's applied bias voltage, respectively. So that, surface potential increases continuously with $V_a$ and the depletion width of the n-Si under graphene widens continuously as the reverse bias increases. However, for the second heterojunction (GIS), the surface potential shows a lower curvature than in the Schottky diode due to the voltage drop through the insulator. Moreover, when a negative bias is applied to the graphene, as in the present case, the energy bands along the whole n-Si substrate bends upwards.

The GIS system behaves as a normal MOS structure, and holes are attracted to the $SiO_2$/Si interface when a negative bias is applied. This in turn can lead to the inversion of the n-Si underneath the oxide.

The condition for the creation of the inversion layer is that the number of minority carriers (holes in this case) at the surface is larger than that of the majority carriers in the bulk (electrons) [43]. When this condition is fulfilled, a considerable difference in the band alignment between the GIS and G/Si junctions along the n-Si substrate is produced. An additional aspect to consider is that the depletion width in the GIS region becomes almost pinned once the inversion layer is created, as the substrate depletion charge is screened by the holes located at the interface [43,44]. The calculations of electron and hole concentrations at the $SiO_2$/Si interface, for two devices with 40 nm and 85 nm oxide thickness, clearly show a rise in hole concentration with increasing $V_R$ (Fig. S3). Furthermore, the thinner insulator results in a lower threshold voltage and consequently in a higher hole density in the inversion layer for the same $V_R$ value. The calculated threshold voltage for 85 nm thick oxide is around -1.2 V, which is in excellent agreement with the bias where a kink is observed in the I-V curves measured in the $G/SiO_2$ region only (Fig. 3h).

All these physical phenomena have been thoroughly studied and represented in Fig. 4. Fig. 4a and 4b show the cross-section of the G/n-Si heterojunction diode with the corresponding simulated plot of the valence band ($E_V$) along the n-Si substrate, exactly at the interface with graphene and SiO$_2$ at V$_R$ = -1 V and V$_R$ = -2 V, respectively, with -2 V ≤ V$_T$ < -1 V, where, V$_T$ is the threshold voltage of the G-SiO$_2$-n-Si junction. Below the G/Si junction, E$_V$ is shifted upwards an amount equal to the applied reverse bias. However, this is not the case for the GIS junction where E$_V$ shows a much modest growth due to the creation of the inversion layer for |V$_R$| > |V$_T$| (Fig 4c). This different behavior gives rise to a high lateral energy gradient that favors hole collection from GIS to G/Si junction.

When our device is illuminated, incident light is absorbed in the Si substrate and electron-hole pairs are generated as a result. It should be noted that graphene has a low absorption coefficient and that the photogenerated carriers have a very short lifetime (in the range of picoseconds [45]). Besides that, SiO$_2$ is transparent for the energy range of the photons we used for the measurements, i.e. 532 nm wavelength laser. Considering these two factors, we can say that graphene and SiO$_2$ do not act as main absorbers in our study, and the same has been observed through SR measurements where the maximum absorption is seen in Si in the visible range (Fig. 1f). Therefore, the photons absorbed in the depletion width of G/Si region or at a distance below the diffusion length of its border produce charge-carriers that are rapidly separated (this is also apparent from Fig. S2, where we observe an increasing photocurrent near the G/Si region). The schematic cross-section shown in Fig. 4e, 4f and the band diagram in Fig. 1e depict that the photogenerated holes are attracted to the surface and electrons go to the Si substrate. The holes that reach the G/Si interface are free to move into the graphene contact. However, in the case of the GIS structure, a thick barrier (here 85 nm thick SiO$_2$) exists that prevents the tunneling and the transit of the photogenerated holes to the contacts. Therefore, the photogenerated holes will start accumulating below the SiO$_2$ as they have to reach the G/Si interface before they can be extracted to the contact. For $|V_R| < |V_T|$, we observed a lower photocurrent for the laser spot located in the GIS region compared to the G/Si region (see Fig. 2b). For these low voltages, the GIS structure is either in the depletion or in the weak inversion region (Fig. 4c and 4e). In either case, photogenerated holes have to travel a long distance close to the Si/SiO$_2$ interface, where the probability of recombination is quite high due to the presence of interface states. If we continue to increase the reverse bias, we finally reach $|V_T|$ and the GIS junction achieves the strong inversion condition. In that case, a high concentration of holes is located at the SiO$_2$/Si interface generating a quasi p-n junction. Therefore, for biases higher than $|V_T|$, the photocurrent measured at GIS increases rapidly, as it is shown in Fig. 2e. For a constant laser power and a fixed depletion width in the GIS region, we do not expect a rise in the generation rate of charge carriers.

Therefore, to explain the increase in the measured photocurrent, we need to consider other factors. First, the larger band bending (surface potential) along the n-Si substrate, as shown in Fig. 4b and 4d, enhances the lateral drift of photogenerated holes in the n-Si substrate from the GIS to the G/Si region following the top of the valence band. The second factor is the formation of the inversion layer, which produces an effective passivation layer for the surface states located at the Si/SiO$_2$ interface [46–48]. As a consequence, a noticeable reduction of surface recombination is achieved and the inversion layer provides a highly conductive path for the minority charge carriers (holes in the present case) [46]. Green *et al.* [46,47], in their pioneering studies of MIS-type photovoltaic cells, have shown that when an inversion layer is formed in Si underneath the oxide, it results in an enhanced solar cell efficiency. Moreover, in our case, the difference in the band alignment between both junctions assists the lateral drift of the accumulated holes in the inversion layer into the G/Si junction and their extraction to the external contacts (Fig. 4d and 4f). Also, the formation of a native oxide layer at the G/Si junction cannot be avoided completely during the fabrication. This may further suppress the photocurrent due to increased recombination processes at the G/Si junction. All these factors result in a higher photocurrent in the GIS region of the photodetector, compared to the G/Si junction, as observed in the present study.

The creation of this type of quasi p-n junctions is a process usually employed to improve the performance of photosensitive devices, as it was shown in the case of organic/Si heterojunctions [49,50]. Yu *et al.* [50] observed the formation of a strong inversion layer near the Si surface in organic-Si nanowire hybrid solar cells, which converts the Schottky contact into a p-n junction resulting in a highly efficient cell. The same effect has recently been proposed for graphene/silicon photodiodes [41,42]. However, in the present case, we not only induce an inversion layer in Si underneath thick $SiO_2$ at reverse biases as low as -2 V, using atomically thin graphene, but also visualize it in present SPC measurements in graphene based photodiodes. This can be attributed to the atomic thinness and transparency of graphene acting as a contact material, which enables the observation of photocurrent generation underneath it.

As is well known from metal oxide semiconductor field effect transistors, the formation of the inversion layer is a bias dependent phenomenon and a stronger inversion layer is formed when thinner insulators are employed [43]. In order to confirm this experimentally, we fabricated G/n-Si photodiodes with two different $SiO_2$ thicknesses of 85 nm and 40 nm. Fig. 5a and 5b show J-V characteristics under white light illumination of both photodiodes in semi-logarithmic and linear scale, respectively. We note a slight saturation in dark current and a notable photocurrent in these diodes in the forward direction. These discrepancies may have their origins in the low contact quality as revealed by the I-V characteristics of the metal-Si contacts (Fig. S3). Nevertheless, below the threshold, the two photodetectors are dominated by the (identical) G/n-Si Schottky diode regions and behave very similarly. However, a distinctively larger photocurrent is observed for the diode with 40 nm thick oxide once the threshold voltage for the inversion layer is reached, because the carrier density in the gated region of this diode is higher for a given $V_R$ (compare simulation results in Fig. S3). The present findings clearly demonstrate that the gated region of G/Si photodiodes contribute significantly to measured photocurrents in such hybrid photodiodes.

In conclusion, we have revealed through scanning photocurrent measurements that the photocurrent in graphene-based hybrid photodiodes is generated not only in the G-Si Schottky barrier region, but also very efficiently in the adjacent G-SiO$_2$-Si (GIS) region, where it strongly depends on the applied reverse bias voltage. It is found that the photocurrent rises sharply by about one order of magnitude in the GIS region above a certain threshold bias voltage, regardless of the incident laser power. We also investigated the effect of oxide thickness on the photocurrent and observed a larger photocurrent for the device with thinner oxide. The observations are explained through simulations by the formation of an inversion layer in Si under the SiO$_2$. This inversion layer not only provides a low resistance path for the minority charge carries, but also acts as a passivation for the surface states in SiO$_2$, thereby enhancing the photocurrent by efficient collection of the charge carriers. The present findings establish the fundamental mechanisms of photocurrent generation in graphene based hybrid optoelectronic devices and may also provide guidelines for designing hybrid photodetectors based on the combination of two- and three-dimensional materials.

## Acknowledgements

We would like to thank Dr. Chanyoung Yim (University of Siegen) for fruitful discussions. Funding from the European Research Council (ERC, InteGraDe, 3017311), the German Research Foundation (DFG LE 2440/1-2 and GRK 1564), European regional funds (HEA2D, EFRE-0800149) and the Spanish Ministry of Education, Culture and Sport (Salvador de Madariaga Program, PRX16/00205) is gratefully acknowledged.

**Methods**

*Device Fabrication:*

A lightly doped n-Si wafer with a thermally grown silicon dioxide (SiO$_2$) layer of 85 nm was used as a substrate. The n-Si wafers were phosphorus-doped with a doping concentration of 2×10$^{15}$ cm$^{-3}$. For chip fabrication, the wafer was diced into 13×13 mm$^2$ samples. Eight photodiodes were fabricated on each chip. The oxide was etched with buffered oxide etchant (BOE) after a first standard UV-photolithography step in order to expose the n-Si substrate. The contact metal electrodes were defined by a second photolithography step followed by sputtering of 20 nm of chromium (Cr) and 80 nm of gold (Au) and liftoff process. The metal electrodes were deposited immediately after the native oxide removal ensuring to form good ohmic contacts. Large-area graphene was grown on a copper foil in a NanoCVD (Moorfield, UK) rapid thermal processing tool. To transfer graphene films onto pre-patterned substrates, ~1 cm$^2$ pieces of graphene-coated Cu foil were spin-coated with Poly methyl methacrylate (PMMA) and baked on a hot plate at 85$^o$ C for 5 minutes. Electrochemical delamination has been used to remove the polymer-supported graphene films from the copper surface [14]. In order to make a good electrical contact between graphene and n-Si substrate, the native silicon oxide on the n-Si substrates was removed by BOE prior to the graphene transfer. Afterwards, the devices were thoroughly immersed into acetone for 3 hours, followed by cleaning them with isopropanol and DI water and drying. At the end, a last photolithography step was performed followed by oxygen plasma etching of graphene in order to define graphene junction areas.

*Electrical Characterization:*

Electrical measurements on the diodes were made with a Karl Süss probe station connected to a Keithley semiconductor analyzer (SCS4200) under ambient condition. The voltage for all devices was swept from 0 to +3 V for forward ($V_F$) and from 0 to -3 V for reverse ($V_R$) biasing. A white light source (50 W halogen lamp) with a dimmer to control the light intensity was used to quickly check that the fabricated diodes are working properly and that they are generally sensitive to light. The intensity of the light source was measured by A CA 2 laboratory thermopile.

*Optical Characterization:*

The spectral response (SR) of the photodetectors was measured using a lock-in technique by comparing it to the calibrated reference detectors. A tungsten-halogen lamp (wavelength ranging between 300 nm and 2200 nm) was used as a light source. Specific wavelengths were selected by a monochromator. The intensity of the light beam was modulated by a chopper with a frequency of 17 Hz. Calibrated Si and Indium-Gallium Arsenide (InGaAs) diodes were used as reference detectors. The photodetectors' currents were measured by pre-amplifiers (FEMTO) and lock-in amplifiers at chopper frequency of 17 Hz for detection of ultra-low currents down to 10 pA. For the responsivity calculation, the measurement principle allows to establish a wavelength dependent correction factor. This correction factor takes into account variations of the preamplifiers, varying photo flux densities caused by the monochromator as well as the area difference between the reference detector and the sample.

*Scanning photocurrent measurements:*

Scanning photocurrent measurements were performed using a Witec Alpha300 R confocal microscope equipped with a piezoelectric scanning stage. The microscope was coupled to a 532 nm wavelength to generate spatially resolved photocurrent, which is converted into a voltage signal using a current preamplifier and is recorded by a lock-in amplifier. The samples to be investigated were mounted on a custom-made sample holder on a PCB.

*Simulations:*

For the simulation of the G/Si Schottky diode and the GIS junction we have solved the 1-D Poisson equation considering a Si substrate with an n-type doping of $2\times10^{15}$ cm$^{-3}$ and a 85nm thermally grown SiO$_2$ layer on top of it. The use of a 1D model is justified by the dimensions of the device under study, which is of the order of hundreds of μm. In the case of the G/Si junction, the extension of the depletion region is below 2μm for $V_R$ = -2 V and even smaller for the G/SiO$_2$/Si junction. For MIS structures, the gradual-channel approximation is typically employed [43], and the 1D-Poisson equation provides a good description of its electrostatic behavior. We have considered that the initial graphene is p-type with an estimated carrier concentration of $p_0$ = 3.5 x 10$^{12}$ cm$^{-2}$. The origin of this charge is thought to be due to charge puddles produced during the graphene transfer onto SiO$_2$ [51]. This intrinsic doping shifts the Fermi level ($E_F$) below the Dirac point ($E_D$) a value

$$E_{F0} - E_D = -\hbar v_F \sqrt{\pi p_0}$$

where $\hbar$ is the reduced Planck's constant and $v_F$ = 1.1 x 10$^6$ m/s is the Fermi velocity of graphene. The values obtained from this calculation provide a good agreement with the experimental ones, such as the Schottky Barrier Height. The charges located in the semiconductor, both, depletion and inversion, are positive as it corresponds to ionized donor impurities and holes, respectively. For the case of the G/Si Schottky diode, we consider only the depletion charge and for the GIS junction we also include the inversion charge.

An equal and opposite charge is induced in the graphene layer, causing an additional Fermi level shift relative to the Dirac point as a function of the applied reverse bias. This additional shift will be opposite to the one produced by the intrinsic charge in the graphene and it can be calculated once the charge in the semiconductor is estimated as a function of the potential in the substrate. Thus, a self-consistent calculation [52] is mandatory and it will provide us information about the depletion width, surface potential, depletion and inversion charge, and conduction and valence band values as a function of the applied bias.

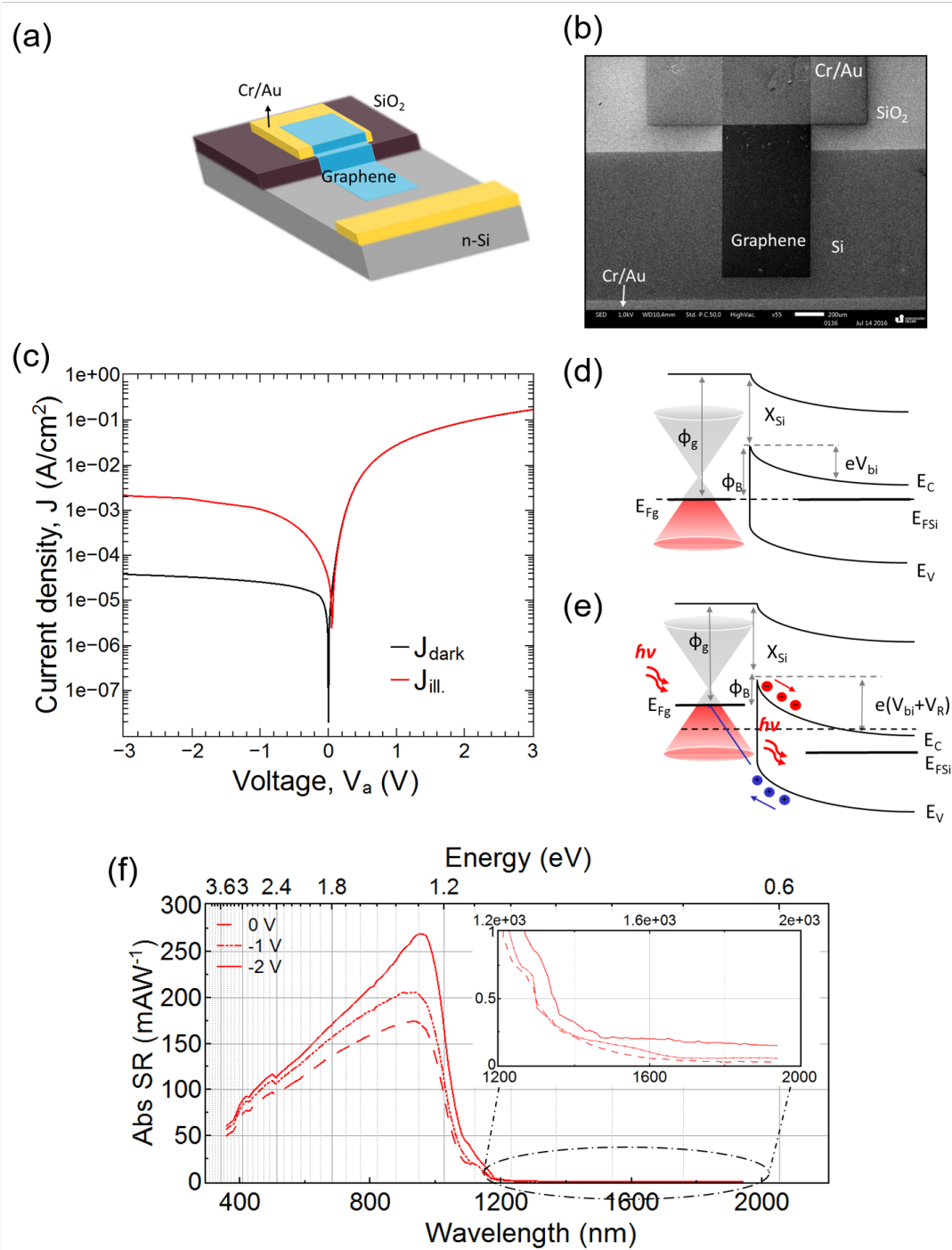

Fig. 1: (a) Schematic, (b) scanning electron micrograph and (c) J-V plot of a graphene – n-Si (G-Si) photodiode in the dark and under illumination. Schematic band diagram of the graphene – n-Si interface (d) in the dark at zero bias voltage and (e) under illumination in reverse biased condition. $E_C$, $E_V$, $E_{FSi}$, $E_{Fg}$, $\phi_g$, $X_{Si}$, $V_{bi}$, $\phi_B$, $V_a$ and $V_R$ indicate conduction band, valence band, Fermi level of Si, Fermi level of graphene, graphene work function, Si electron affinity, built-in potential, Schottky barrier height (SBH), applied voltage and reverse bias voltage of the diode, respectively. (f) Abs. SR vs. wavelength (lower x-axis) and energy (upper x-axis) of the graphene – n-Si photodiode for wavelengths ranging from 360 nm (3.44 eV) to 2200 nm (0.56 eV) and the inset shows zoom-in from 1200 nm (1.03 eV) to 2000 nm (0.62 eV) at zero bias and reverse biases of -1 V and -2 V.

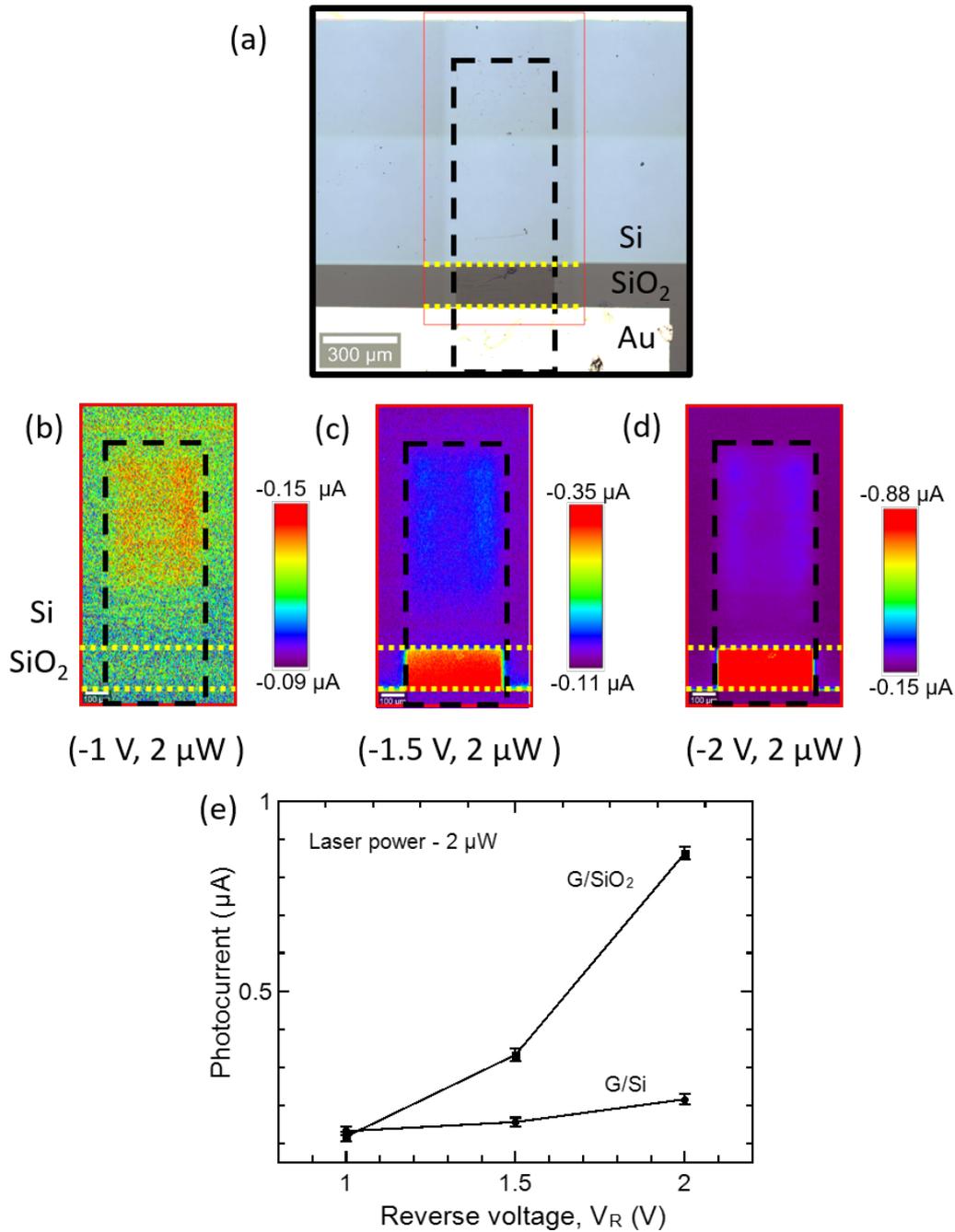

Fig. 2: Scanning photocurrent (SPC) measurements of the diode at various reverse biases. (a) Optical micrograph of the diode. The area inside the red rectangle was scanned for photocurrent measurements. The graphene region is represented by the black dashed rectangle. The horizontal yellow dashed lines inside the scanned area depict the $SiO_2$ region and the same is valid for all the images. Photocurrent maps of the scanned area at a laser power of 2 μW and a reverse bias of (b) -1 V, (c) -1.5 V and (d) -2 V. (e) Evolution of the photocurrent (shown as absolute values) with increase in reverse voltage in G/Si (circle symbol) and $G/SiO_2$ (rectangle symbol) regions at a laser power of 2 μW. A higher current is observed in $G/SiO_2$ region compared to G/Si region at reverse biases slightly above 1 V.

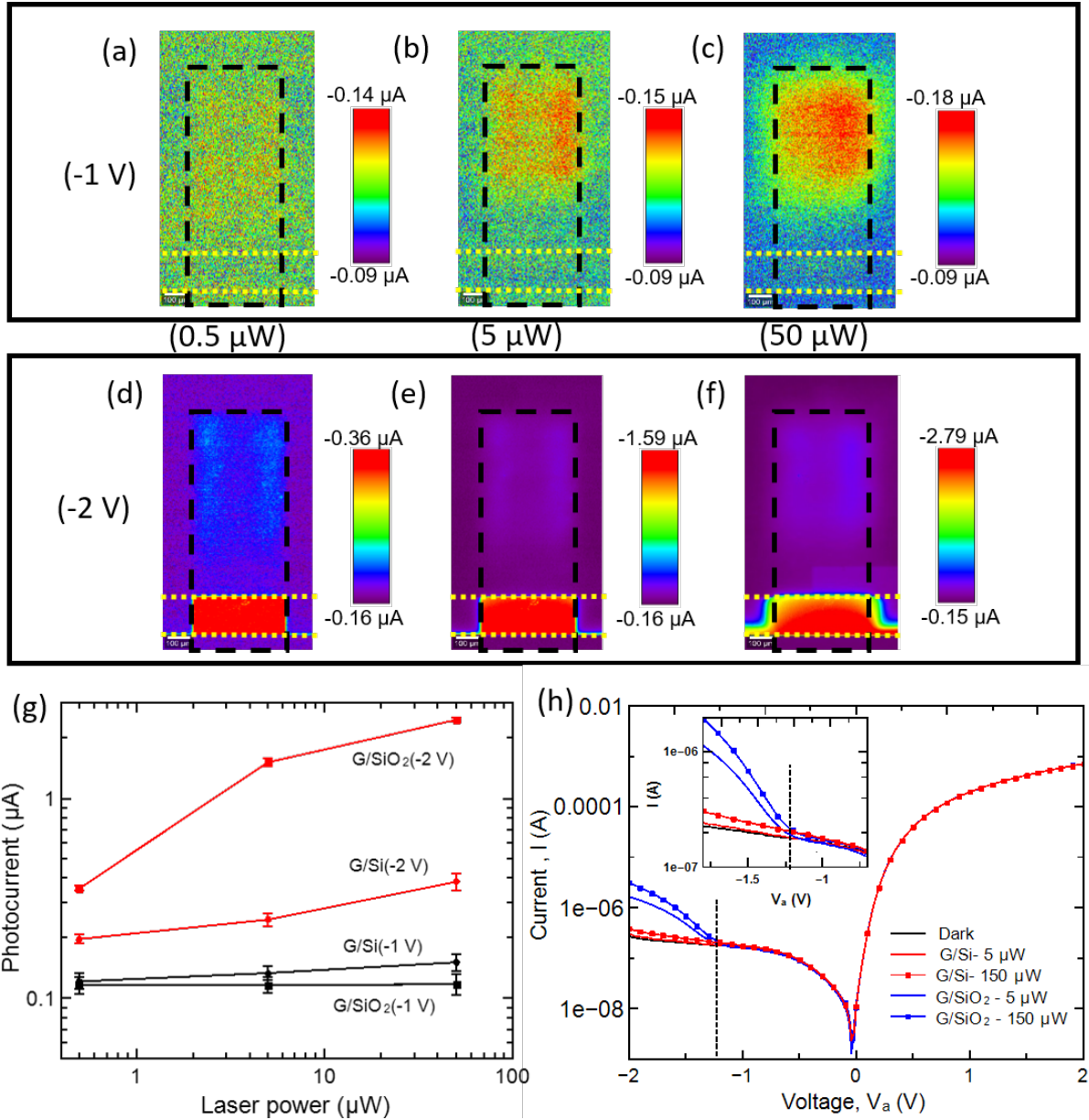

Fig. 3: Laser power dependence of photocurrent. Photocurrent maps of the diode at a reverse bias of -1 V with a laser power of (a) 0.5 µW, (b) 5 µW, and (c) 50 µW. Photocurrent maps of the diode at a reverse bias of -2 V with a laser power of (d) 0.5 µW, (e) 5 µW, and (f) 50 µW. The graphene region is represented by the black dashed rectangle. The horizontal yellow dotted lines depict the SiO$_2$ region. (g) Evolution of the photocurrent at different laser powers of 0.5 µW, 5 µW and 50 µW and at reverse biases of -1 V and -2 V in G/Si (circle symbols) and G/SiO$_2$ (rectangle symbols) regions. A significant photocurrent is observed at the Au contact periphery at a bias of -2 V, which increases for higher laser power. Also, it is found that region of higher current is independent of laser power, rather it depends on the applied bias. (h) Current-voltage (I-V) curves obtained for G/SiO$_2$ and G/Si regions of the diode under dark and laser illumination at two different laser powers i.e. 5 µW and 150 µW. A kink is clearly observed at an applied reverse bias of around 1.2 – 1.3 V. The inset in (h) shows the magnified curves in the bias range from -0.7 V to -1.8 V to emphasize the kink formation at a particular bias. The vertical dashed lines mark the position of the threshold voltage at which the current starts rising sharply.

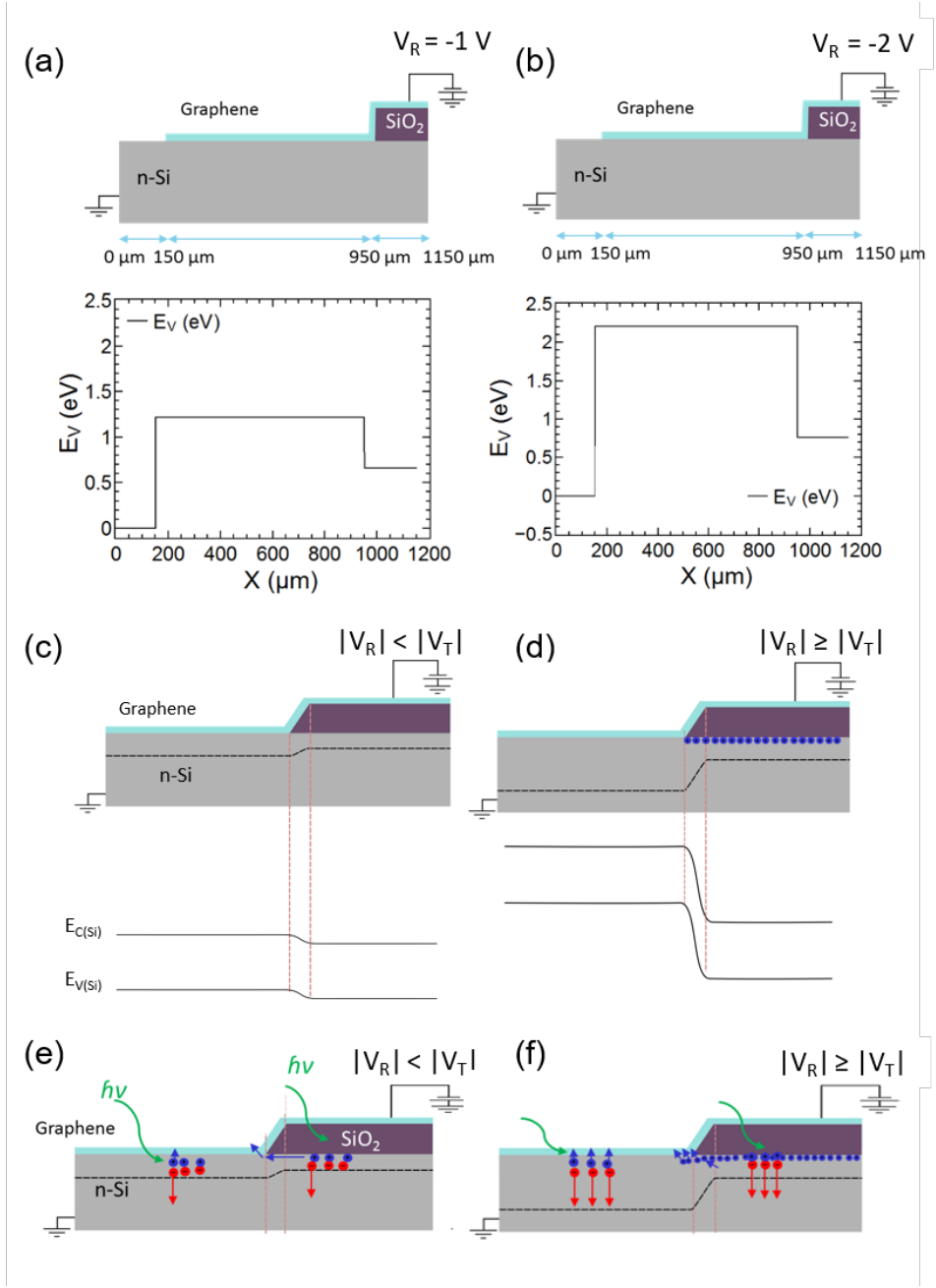

Fig. 4: Cross-section of the graphene – n-Si heterojunction diode with corresponding simulated plot of the Valence Band ($E_V$) along the n-Silicon, just at the top interface with graphene and $SiO_2$ in reverse biased condition of (a) $V_R$ = -1 V ($|V_R| < |V_T|$) and (b) $V_R$ = -2 V ($|V_R| \geq |V_T|$ ) in the dark. The conduction band ($E_C$) will be parallel to the $E_V$ and shifted by the Si bandgap. Schematics, which show (c) formation of depletion layer in n-Si under graphene and G/$SiO_2$ at $|V_R| < |V_T|$ and (d) widening of depletion width in n-Si under graphene and formation of inversion layer in n-Si under $SiO_2$ at $|V_R| \geq |V_T|$ in the dark and (e) and (f) under illumination. $V_R$, $V_T$, $E_{C(Si)}$ and $E_{V(Si)}$ indicate reverse voltage, threshold voltage, conduction band and valence band of n-Si, respectively. Plotted dashed lines in n-Si region are representative of depletion layer. The formation of an inversion layer in Si, underneath G/$SiO_2$ region, above a threshold reverse voltage results in a higher photocurrent in that region. The holes in the inversion layer may fill the trap states at the $SiO_2$/Si interface, therefore, allowing more efficient collection of photogenerated holes. This results in a higher photocurrent with increasing reverse bias.

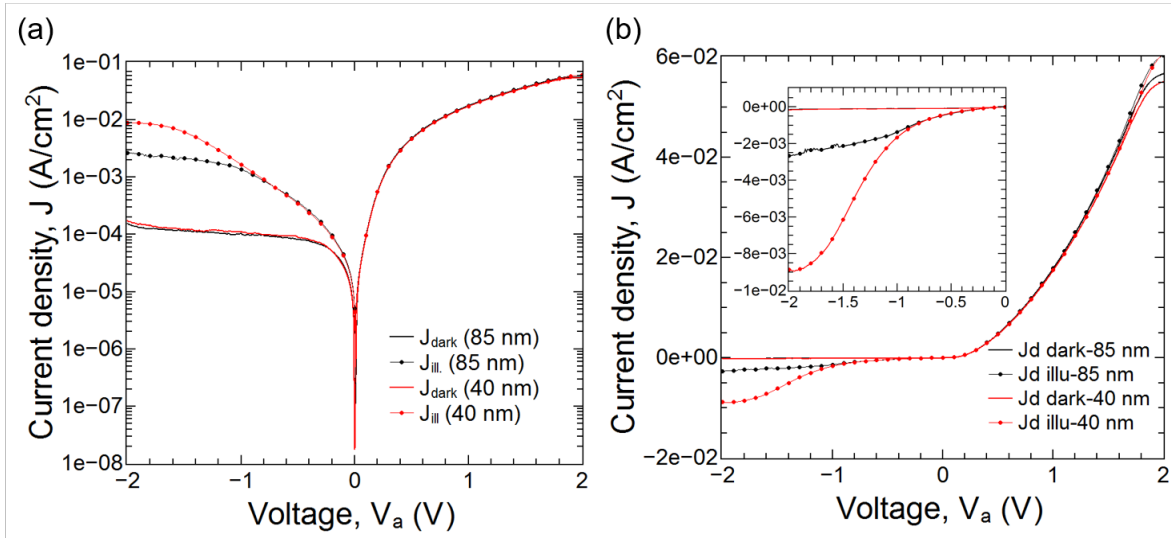

Fig. 5: J-V comparison of graphene/n-Si photodiode with two different SiO$_2$ thicknesses of 85 nm and 40 nm on (a) semi-logarithmic and (b) linear scale in the dark and under illumination. The inset shows zoom-in for reverse bias voltage.